# Velocity distribution of larger meteoroids and small asteroids impacting Earth

Esther Drolshagen[1*], Theresa Ott[1*], Detlef Koschny[2,3], Gerhard Drolshagen[1], Anna Kristiane Schmidt[1], and Björn Poppe[1]

[1] Division for Medical Radiation Physics and Space Environment, University of Oldenburg, Germany,

[2] ESA/ESTEC, Noordwijk, The Netherlands,

[3] Chair of Astronautics, TU Munich, Germany.

*These authors contributed equally to this work

Esther.Drolshagen@uni-oldenburg.de; Theresa.Ott@uni-oldenburg.de

**Keywords**: Meteoroids, Meteors, Near-Earth object, Earth impacting objects, Meteoroid and asteroid velocity distribution, Unbiased velocity distribution

## Abstract

Various meteor and fireball networks exist worldwide. Most data sets which include ground-based observational data of meteors are affected by biases. The larger and faster the entering meteoroid, the brighter is the produced meteor. Hence, small and slow objects often stay undetected. This bias of meteor observations towards faster meteoroids is a challenge if quantitative population and flux models are derived. In this work the velocity distribution of objects in space is analysed by using different data sets that are not affected by this velocity bias since they include only large objects, like the near-Earth object (NEO) risk list of ESA's (European Space Agency) SSA (Space Situational Awareness) near-Earth object Coordination Centre (NEOCC), and the fireballs in NASA's (National Aeronautics and Space Administration) CNEOS (Center for near-Earth object Studies) JPL (Jet Propulsion Laboratory) fireball database. Additionally, when only the largest objects recorded with the CILBO (Canary Island Long-Baseline Observatory) camera setup were analysed, a very similar distribution was shown. These velocity distributions are in good agreement with a widely used velocity distribution for smaller sporadic meteoroids in free space which was adopted as reference by the ECSS (European Cooperation for Space Standardisation) Space Environment Standard.

1. **Introduction**

According to Drolshagen *et al.* (2017) about 54 tons of extra-terrestrial material fall down on Earth every day. Other models estimate daily accumulation masses of Earth between 5 t and 300 t (e.g. Ceplecha, 1992 and references therein, Brown *et al.,* 2002). At least values between 30 t - 150 t are within the range of uncertainty of the latest measurements and models. For a more detailed discussion see Drolshagen *et al*. (2017). The impacting material is interplanetary dust, meteoroids, or asteroids. The largest mass influx is from meteoroids in the size range of some tens of microns to about 1 mm.

The flux, velocity, and directional distribution of objects in space and impacting Earth provides information on the sources of these objects. This information is also required to develop flux models for a quantitative impact risk assessment of spacecraft in orbit. Several meteoroid engineering models have been developed for Earth orbit (Smith, 1994; Divine, 1993; Staubach *et al.,* 1997; McNamara *et al.,* 2004; Dikarev *et al*., 2005) and others for interplanetary space (ibid.). Recently, meteoroid flux models have also been developed for special locations such as for the local Jovian system (Liu *et al*., 2016), and for special events such as meteor storms (Cooke and Moser, 2010) or a close encounter of Mars with a cometary coma (Moorhead *et al.*, 2014). The influx of larger objects, starting at a size of a few metres, is relevant for the topic of Planetary Defence, as these may cause damage on our planet.

When a meteoroid or small asteroid enters the Earth's Atmosphere it ablates and excites the surrounding air. This leads to a visible meteor or fireball (a meteor brighter than magnitude -4). The brightness of a meteor or fireball depends mainly on the mass and velocity of the impacting object. Especially, it is a strong function of the velocity. This holds true for radar or optical meteors alike. There are various publications and formulas which relate the mass and velocity of the impactor to resulting meteor brightness, see e.g. Verniani (1973), Ceplecha and McCrosky (1976), Halliday *et al.* (1996), or Weryk and Brown (2013). A common feature of all these formulas is a strong dependence on the impact speed. This strong dependence of the brightness on velocity introduces a clear bias towards meteoroids with higher velocities. This is most obvious for meteor showers which usually have high impact velocities. The enhanced meteor activity during such streams is mainly resulting from the high impact velocity which makes smaller, and more abundant, objects visible as meteors. At lower velocities the meteoroid streams would be hardly noticeable within the sporadic background population.

Velocities of meteoroids entering the Earth's Atmosphere can range from about 11 km/s to above 73 km/s. The gravitational attraction of Earth determines the lower limit of this range, particles on interplanetary trajectories hit Earth with speeds exceeding the escape velocity (approximately 11.1 km/s at a typical meteor altitude of 100 km). The upper limit of about 73.6 km/s is a combination of the Earth's maximum orbital velocity around the Sun, the speed of an object on a parabolic heliocentric orbit at the corresponding point in the Earth's orbit, and the Earth's gravitational pull (this again corresponds to an altitude of 100 km). Only particles hitting Earth "head-on" can achieve such velocities. For Earth-orbiting spacecraft the impact speed range can be extended from about 4 km/s to 81 km/s because of the spacecraft orbital motion.

There is no minimum impact velocity in interplanetary space. An average or characteristic impact velocity around 20 km/s is sometimes assumed for 1 au (e.g. Grün *et al*., 1985) but impact velocities are better described by a distribution of velocities. The velocity distribution differs for the existing meteoroid models. Examples are the Meteoroid Engineering Model (McNamara *et al*., 2004); or the alternative model from Taylor (1995). Different speed distributions of several models are discussed in Drolshagen *et al.* (2008). Impact velocities will tend to be higher near the Sun and lower at larger distances from the Sun (see e.g. Divine, 1993; Staubach *et al.,* 1997; Dikarev *et al*., 2005).

Stream meteoroids usually have higher impact velocities compared to the ´typical´ speed of 20 km/s (at 1 au) for sporadic meteoroids. For instance, the Perseids intercept the Earth at fixed speeds of nearly 60 km/s, the Leonids at 71 km/s, and the Geminids at 35 km/s (Jenniskens, 1994; SonotaCo, 2009; Brown *et al*., 2010; Jenniskens *et al*., 2016). Depending on the reference used, the reported stream velocity can differ by 1 km/s - 2 km/s.

The bias of meteor observations towards faster meteoroids is a problem for the derivation of quantitative population and flux models. Small meteoroids are only detected if they are fast enough. It is assumed that in most data sets slow meteors are under-represented. In contrast, larger objects can also be detected if they are slow. Reference velocity distributions for meteoroids which are based on meteor observations have to correct for the velocity bias (Taylor, 1995). Any such bias correction still contains large uncertainties. Looking only at larger objects is one possibility to obtain the un-biased velocity distribution of interplanetary material, at least for these larger objects. This approach will be followed here by studying the velocity distribution of larger meteoroids and small asteroids. Recent observations have provided sufficient data for these larger objects.

In literature diverse studies were performed to correct for this bias and different bias correction methods were applied. Since raw brightness-limited (by optical means) or electron-line density limited (by radar) surveys have strong speed biases the final distribution is heavily dependent on assumptions of the mass distribution and luminous or ionization efficiency etc. Nonetheless, most works that corrected for this bias also found that most of the meteoroid population near the Earth has low velocities (< 20 km/s). We will only give a brief overview of previous works.

In Erickson (1968) the velocity distribution of 286 photographic meteors is corrected. Using only the brightest meteors (brighter than 1 mag) they found an average velocity of 16.5 km/s and presented a velocity distribution of non-varying mass with a steep increase towards smaller velocities and a maximum around 14 km/s.

A detailed analysis was carried out for the meteors detected with the Harvard Radio Meteor Project (HRMP) and continually improved in follow up papers, see Sekanina and Southworth (1975), Taylor (1995), and Taylor and Elford (1998). The most complete HRMP distribution is presented in Taylor and Elford (1998), see their Fig. 3. They correct for "[…] the observing schedule, antenna beam patterns, the radio diffusion ceiling, speed dependence of ionization production, the flux enhancement due to the Earth's gravity and the probability of encounter with the Earth." (Taylor and Elford, 1998). They concluded with a velocity distribution including mainly meteors with low velocities. Their distribution peaks at around 13 km/s, slowly decreases until velocities of ca. 45 km/s from which point on it rises again. As it can be seen in their Fig. 3 their velocity distribution shows large deviations from the one published in the previous work by Taylor (1995). This clearly demonstrates the large uncertainties based on assumptions made and on effects for which a correction is already applied.

Hunt *et al.* (2004) applied the model of Taylor (1995) to radar meteoroid velocity distributions derived from data collected by the Advanced Research Project Agency Long Range Tracking and Instrumentation

Radar and by the HRMP. Their corrected, mass-weighted, distributions are quite similar and consistently show clear peaks around 20 km/s, see their Fig. 9.

Galligan and Baggaley (2004) analysed radar meteors detected by the Advanced Meteor Orbit Radar with a focus on debiasing. Their resulting orbital distributions are quite similar compared to the ones published by Taylor and Elford (1998). However, if the two velocity distributions are directly compared, as presented in Fig. 34 of Galligan and Baggaley the velocity distribution of Galligan and Baggaley shows slightly larger velocities. They divide between geocentric and heliocentric speeds (see their Fig. 33) with distributions with peaks at around 30 km/s and 40 km/s, respectively. The complete distribution increases until its maximum at roughly 33 km/s, then decreases until about 50 km/s. It remains relatively constant until it decreases at larger velocities of ca. 60 km/s again. Between 30 km/s and 40 km/s the distribution has a plateau-like shape.

In Janches *et al.* (2003) the geocentric velocity distribution of micrometeors detected with the Arecibo radar is presented. They found different distributions for different pointing directions and four different populations which they characterized by the geocentric velocities of the particles. The largest meteoroid population was found to be the two "fast" (45 km/s and 50 km/s) groups. Even when they analysed possible observation biases, they found a meteoroid population lacking objects at velocities < 30 km/s and hence, a velocity distribution with a very different shape than similar studies.

More recently, in Moorhead *et al.* (2017) a detailed study of the biases of radar meteors collected with the Canadian Meteor Orbit Radar was released. Therein, a review of earlier performed debiasing studies for radar meteors can be found. They found an increase of slow meteors after the debiasing, even larger than in previous works. Their presented debiased velocity distributions show maxima at around 17 km/s (compare their Fig. 13) varying with the assumptions made during the process.

Williams *et al.* (2017) published a study based on meteors detected with the Jicamarca high-power large-aperture radar. They applied different velocity corrections on the data with a detailed evaluation of how the assumed ionization coefficient for incoming meteors in the velocity range between 10 – 20 km/s effects the result. They utilized previously published values for this coefficient. From their corrected distributions they conclude a monotonic increase of meteoroid numbers with decreasing velocity and mean velocities between 20.5 km/s and 16.5 km/s, compare their Fig. 2.

Le Feuvre and Wieczorek (2011) discussed an orbit and size distribution model for asteroids based on previous publications. They use it (amongst others) to model the cratering of the Moon. As it can be seen in their Fig. 4 the impact velocities of asteroids with the highest impact probability on Earth are in the lower velocity range. The velocity distribution has a peak around 15 km/s, a steep increase for slower velocities, and a rather shallow decrease towards velocities around 40 km/s. They assumed a mean impact velocity of 20 km/s based on the model of Brown *et al.* (2002).

Greenstreet *et al.* (2012) presented a NEO orbital distribution model and a corresponding impact speed distribution, see their Fig. 10. The distribution shows that most objects have smaller velocities with a sharp peak around 15 km/s and a long tail for faster velocities up to around 45 km/s. Their mean impact speed is 20.6 km/s.

Chesley and Spahr (2004) analyzed a large set of simulated asteroids based on the known NEO population with focus on surveying and monitoring. They presented a velocity distribution for their synthetic impactors. About 30 % of their impactors are slower than 17 km/s, 50 % have a velocity lower than 20 km/s, and almost 70 % are slower than 25 km/s, see their Fig. 2.2.

Rumpf *et al.* (2017) analyzed "Asteroid impact effects and their immediate hazards for human populations". To do so, they used a set of impactors published by Chesley and Spahr (2004). In Rumpf *et al.* the impact speed distribution of the impactors can be found in their Fig. SII d with a lognormal distribution fit. This figure shows that the data is described well when compared to a probability density function of a logarithmic normal distribution. They found a mean of the lognormal distribution of 16.6 km/s. In this work different velocity distributions of data sets that should not be affected by the velocity bias are compared. First, there are three data sets introduced and the velocity distributions computed. The first set (Section 2.1) is the fireball database from NASA's (National Aeronautics and Space Administration) CNEOS (Center for near-Earth object Studies) at JPL (Jet Propulsion Laboratory). This database contains information on objects which have impacted Earth since 1988 with an impact energy of 0.1 kt TNT equivalent or larger (CNEOS, 2018). The second data set (Section 2.2) are all the NEAs (near-Earth asteroids) that are listed on ESA's (European Space Agency) SSA (Space Situational Awareness) risk list, the NEODyS Virtual Impactors (VI) (SSA, 2018). This risk list contains all known asteroids which have a chance to hit Earth within the next 100 years. Potential impact dates and velocities are obtained by Monte Carlo simulations sampling the region of orbital uncertainty, see e.g. Farnocchia *et al.* (2015). The displayed data refer to the VI with the largest Palermo Scale value (SSA, 2018). Finally, the velocity distribution of the largest meteoroids (> 1 g) for which the corresponding meteors were detected with video cameras of the CILBO (Canary Island Long-Baseline Observatory) system of ESA is analysed (Section 2.3). Earlier analyses of subsets of the CILBO data (Drolshagen *et al.*, 2014 and Koschny *et al.*, 2017) have shown that all objects larger than 1 g should have been detectable by the cameras, independent of their velocities. It is believed that all three of these data sets are free of any velocity bias and show the true velocity distribution of these objects if they hit Earth. As a reference, the velocity distribution published by the ECSS (European Cooperation for Space Standardization (ECSS, 2008)) adapted to 100 km above Earth's surface is described (Section 2.4). This distribution is based on the work by Taylor (1995). For comparison the velocities of the major meteor showers are presented (Section 2.5) as well.

2. **Velocity distributions**

In the following sections the data sets discussed in this work will be described and their velocity distributions compared starting with the JPL fireball database, followed by the objects of the SSA asteroid risk list, and the CILBO data recording of larger meteoroids. The adapted distribution of the ECSS will be used for comparison and some major meteor showers are presented to show their influence on the velocity perception. The main objective of this work is a comparison of velocity distributions of objects which either did hit the atmosphere of Earth or could do so in the future as calculated from their known orbit. All objects considered were either directly observed during impact or are known asteroids. We also fitted analytical expressions to the observed velocity distributions. These fits are presented as additional information and they are not the main goal of the study. It was not our intention to derive the best possible fit or to promote a new velocity distribution. They are a convenient expression which give reasonable fits to the data.

### 2.1. JPL database of fireballs

The fireball and bolide database from NASA's CNEOS at JPL of the California Institute of Technology is based on the data of US government satellites (CNEOS, 2018). It lists world-wide fireball events for objects with large impact energies ($\geq$ 0.1 kt TNT equivalent) and hence, large sizes. The table includes e.g. the information on the fireball event's date, location, and impact energy. For some cases the object's velocity is available as well. It is stated that this database does not include all fireballs detected by this system but is a smaller subset of all available information. The list is continuously updated but not in real-time. There are 767 fireballs online, status as of 29 January 2019, for 183 of them the velocity is given.

Two of these objects have a reported velocity smaller than 11.1 km/s (9.8 km/s (1 July 2008) and 10.9 km/s (19 April 2018)). This can have at least two possible explanations, measurement uncertainties or the entering object is not of natural origin but instead man-made space debris. A re-entering man-made object can have a velocity as low as about 8 km/s if it was orbiting or even lower for ballistic re-entries. A comparison with the ESA re-entry prediction database, ESA's DISCOS (Database and Information System Characterising Objects in Space, DISCOS, 2018) database, did show no possible match for the event which happened on 19 April 2018, the origin of the other event needs further investigation since there are several possible matches in the database (DISCOS, 2018). Additionally, a comparison with the Aerospace database (Aerospace, 2018) did not yield a match for the more recent event. The event from 2008 could not be compared with this database since the time range of the publicly available data is limited. To summarize, the reason for the low velocities of these events needs to be analysed more detailed which goes beyond the scope of this work. At the moment, the two events were interpreted to be caused by natural objects and taken into consideration for the analysis.

For all fireballs with available velocity information the data is binned in 2 km/s bins and normalized. The symbols show the middle of each histogram bin. The result is shown in Figure 1 as orange stars.

The highest measured velocity is 49 km/s, hence the velocities range from about 9 km/s – 49 km/s. The mean value of the velocities is 17.73 km/s and the median is 16.70 km/s. The maximum of the distribution is around 15 km/s.

The accuracy of the published velocity values is not well known. It is stated that the given velocities are for a position just before the location of main fireball brightness, which is at typical altitudes of 25 km - 40 km. Compared to the top of the atmosphere at 100 km altitude the additional acceleration by Earth will only amount to about 0.05 km/s. Deceleration will be caused by atmospheric friction, but for these larger objects the effect should not be very big either.

The fireball database includes two objects which were detected in space before impact, 2008 $TC_3$ and 2018 LA. For these two impactors the orbit is well known and the impact velocity at the top of the atmosphere can be calculated. The impact velocities calculated from the orbit are 12.78 km/s for 2008 $TC_3$ and 16.60 km/s for 2018 LA (Micheli, 2019). This compares to reported fireball velocities of 13.3 km/s (for 2008 TC$_3$) and 16.9 km/s (for 2018 LA). All velocities should be for a geocentric, Earth-fixed coordinate system. This means that the reported measured velocities are 0.3 km/s and 0.52 km/s higher than the impact speeds calculated from the orbit. This is a bit surprising as one might expect some atmospheric deceleration of the fireball. Lacking a better explanation, we assume that the difference is due to measurement uncertainties of the fireball data. For 2008 TC$_3$ some more information on the velocity is provided at the end of Section 2.6. For meter-scale impactors in the JPL data Brown *et al.* (2016) estimated the uncertainty in the velocity to be in the order of 0.1 km/s – 0.2 km/s, whereas Devillepoix *et al.* (2019) found deviations of up to 28 % in velocity from independent observations for two of ten analysed events. An uncertainty of 1 - 2 km/s is a common value used for this database. However, we are aware that the true uncertainty could be larger. The assumed uncertainty of the velocity will not affect the essence of our presented results.

The velocity distribution is presented in Figure 1. The vertical bars give the statistical errors. Various fits were applied to derive an analytical expression for the velocity distribution. Their agreement with the data distribution and fitting stability for different bin sizes were explored. The fits used in this work were carried out with Origin (OriginLab Corporation, Northampton, United States) (OriginLab, 2018) and the data analysis and displaying with a Python script (Python Software Foundation, Beaverton, United States) (Python Software Foundation, 2018). The logarithmically spread normal distribution (lognorm, see OriginLab (2018b)) was found to fit the data quite well over the entire velocity range. Furthermore, this distribution was found to yield comparably stable results for changing bin sizes. Hence, the log-normal distribution is the one used in the following to describe the velocity distributions and presented as a dashed green line in Figure 1.

Following Limpert *et al.* (2001) a log-normal distribution can be described as shown in Eq. (1). Usually, the distributions are described with the log-transformed variable with the mean $\mu$ (or expected value) of the distribution and the standard deviation $\sigma$ as parameters (Limpert *et al.* (2001)).

$$y = \left(\frac{1}{x \cdot \sigma \cdot \sqrt{2 \cdot \pi}}\right) \cdot \exp\left(-\frac{1}{2 \cdot \sigma^2} \cdot (\ln(x) - \mu)^2\right) \tag{1}$$

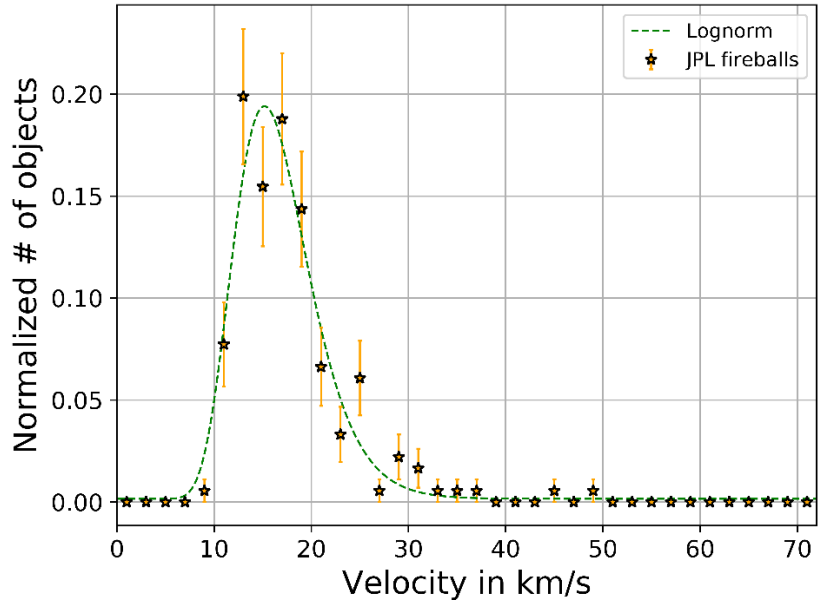

*Figure 1: Velocity distribution of 183 JPL/CNEOS fireballs.*
*The orange stars show the velocity distribution of the objects in the fireball database published by NASA's CNEOS at JPL with error bars, compare Section 2.1. The dashed green line indicates the fit of a lognormal distribution that is applied to the data. No velocities larger than 50 km/s were found in the database.*

The formula with resulting fitting parameters is given in Eq (2). To test the quality of the fit a Kolmogorov–Smirnov test was conducted, to statistically quantify the agreement between the data points (velocities between 9 km/s and 49 km/s) and the fit. The test showed a good agreement between data and fit. With a significance level of 0.01 the data follows the fit and does not differ significantly. The mean value and standard deviation of the distribution are $\mu_{JPL} = 15.31$ km/s and $\sigma_{JPL} = 3.43$ km/s (computed as explained in OriginLab (2018b)).

$$y = 0.0016 + \frac{1.90}{\sqrt{2 \cdot \pi} \cdot 0.25 \cdot x} \cdot exp\left(-\frac{\left(ln\left(\frac{x}{16.17}\right)\right)^2}{2 \cdot 0.25^2}\right) \tag{2}$$

A resulting velocity distribution that is best fitted by a logarithmic normal distribution is in good agreement with the results found by Rumpf *et al.* (2017). Moreover, the overall shape of the velocity distribution presented in Figure 1 is in good agreement with further velocity distributions in literature, like the one of Chesley and Spahr (2004), even though theirs and ours are based on very different data sets. The Chesley and Spahr (2004) impactors are based on simulated impactors based on the known NEO population, the distribution presented in this chapter is based on satellite data of detected impacts.

### 2.2. SSA-NEAs - the NEODyS Virtual Impactors

The risk page of ESA's near-Earth object Coordination Centre (NEOCC) includes all known objects which show a computed impact solution with Earth (Koschny and Drolshagen, 2015). As of 29 January 2019, this list includes 815 objects (SSA, 2018). The SSA NEOCC regularly publishes details about the objects, like their size and velocity as well as the date, and the probability of the potential impact. "Each entry contains details on the Earth approach posing the highest risk of impact (as expressed by the Palermo Scale)" (SSA, 2018). Three of the 815 SSA-NEAs have a date of the potential impact event which is still more than 100 years away (SSA, 2018) and will not be taken into consideration for this analysis. The orbits of the NEAs are determined with a software package called OrbFit (NEODyS, 2018) from astrometric data available at the Minor Planet Center. By Monte Carlo simulations sampling the region of orbital uncertainty the potential impact dates and velocities for so called virtual impactors are obtained. This procedure is called impact monitoring, see *e.g.* Farnocchia *et al.* (2015).

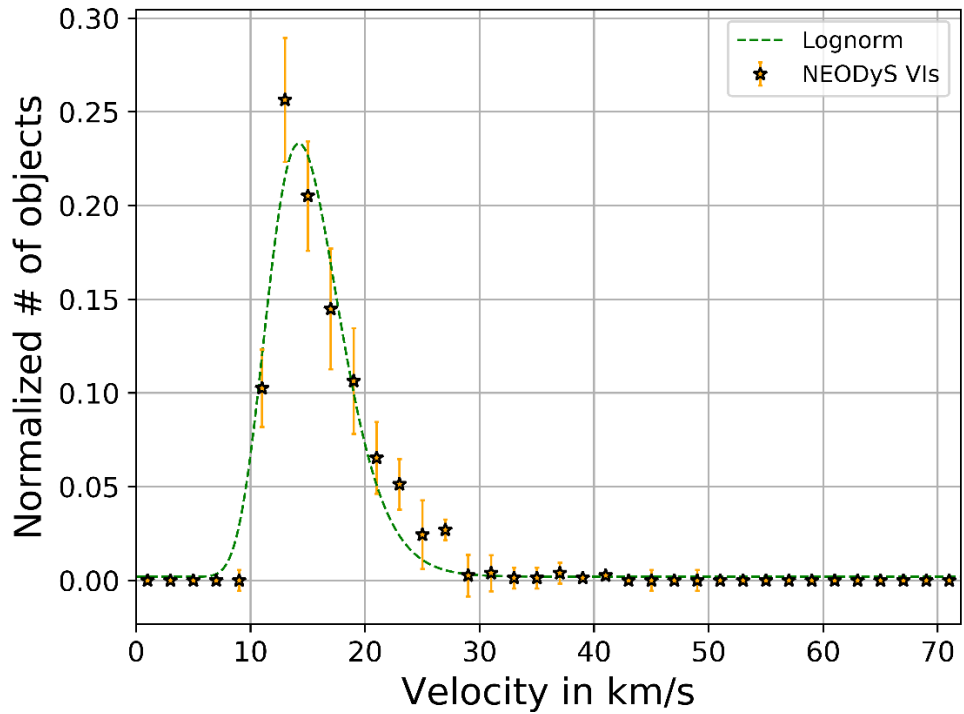

*Figure 2: Velocity distribution of NEODyS Virtual Impactors (NEODyS VIs) with fitted curve.
The orange stars show the normalized velocity distribution of the objects listed in the risk list of ESA's NEOCC with error bars, see Section 2.2. The dashed green line indicates the fit of a lognormal distribution. No velocities larger than 41 km/s were found in the database*

One asteroid can have several virtual impactors which can impact at different times. Information on all these potential impact solutions is available, but only the information of the virtual impactor with the highest impact probability is included in the risk list. For all of these NEODyS Virtual Impactors in the risk list, called VIs in the following, the impact velocity information at the Earth's surface, without atmosphere, is available. The calculated velocities should be accurate to better than 0.1 km/s. While the orbit has usually considerable uncertainties, this region of uncertainty is sampled by many (several thousand) virtual objects. If one of these objects is predicted to become a VI and hit Earth, the orbit and impact velocity of this VI are well determined.

The velocities are extracted, binned in 2 km/s bins, and normalized. The result is shown in Figure 2 as orange stars. The vertical bars give again the statistical errors. The smallest calculated impact velocity is 11.18 km/s, the highest 41.02 km/s, the mean 16.48 km/s and the median 15.33 km/s. The maximum of the distribution is around 14 km/s.

As for the JPL data, the green dashed line in Figure 2 describes the fit of a log-normal distribution. The formula with parameters is given in Eq. (3) and to test the quality of the fit, again, a Kolmogorov–Smirnov test was performed. It was used to obtain a statistical quantification of the agreement between the data points (velocities between 11 km/s – and 41 km/s) and the fit. The data seems to follow the fit and does not deviate significantly with a significance level of 0.01.

$$y = 0.0020 + \frac{1.87}{\sqrt{2 \cdot \pi} \cdot 0.22 \cdot x} \cdot exp\left(-\frac{\left(ln\left(\frac{x}{14.94}\right)\right)^2}{2 \cdot 0.22^2}\right) \tag{3}$$

The mean value and standard deviation of the distribution are $\mu_{\text{NEODyS VIs}} = 15.31$ km/s and $\sigma_{\text{NEODyS VIs}} = 3.43$ km/s. As for the JPL data the expected values obtained from the log-normal fit are about 1 km/s lower than the mean value directly obtained from the data but matches the median values very well.

Furthermore, the velocity distribution of the set of the synthetic impactors published by Chesley and Spahr (2004) is in good agreement with our distribution. For the distribution presented in this chapter, they are based on quite similar data. The Chesley and Spahr (2004) impactors are based on simulated impactors based on the known NEO population while the distribution presented in this chapter is based on virtual impactors of known NEAs. Both distributions have a very similar steep peak around 15 km/s and a long tail towards larger velocities. The mean of the lognormal distribution of the speed distribution of their synthetic impactors is 16.6 km/s as published by Rumpf et al. (2017). This value is only slightly larger than the mean value of 15.3 km/s presented here.

The NEODyS VI data include all known asteroids which could hit Earth within the next 100 years. A given asteroid could have several different impact solutions (i.e. hit Earth at different times and with different impact geometries and impact probabilities). In the present study only the potential impact with the highest impact probability is considered.

It is our aim to obtain the unbiased velocity distribution for these objects. This requires consideration of a potential bias w.r.t. size or velocity of the known near-Earth asteroids. It is believed that less than 1 % of NEOs smaller than 30 m in size are presently known while about 90 % of objects larger than 1 km are catalogued.

As the asteroid detectability is dependent on the object's brightness the database is absolute magnitude (*H*) limited.

Furthermore, smaller objects are only detected if they are rather close to Earth. It could be possible that biases not only favour larger asteroids but also discriminate against objects with high relative velocity, especially those for which a high velocity is due to high inclination which are more challenging to discover. Hence, high velocity impactors may be underrepresented in the VI sample.

To investigate whether the velocity data is biased with size, it will be divided into three subsets, selected by *H*-magnitude. The SSA risk list includes the size of the objects. In most cases it is derived from the absolute magnitude value with Eq. (4), assuming an albedo of 0.1 (SSA, 2019).

$$Size = \frac{1.329 \cdot 10^6}{\sqrt{0.1}} \cdot 10^{-0.2 \cdot H} \qquad (4)$$

For this work the size values are taken from the risk list page and from these the *H*-values are computed following Eq. (4). Only for 13 of the 812 objects *H*-values are available from other sources, for example from infrared observations for (99942) Apophis. For those asteroids the alternative values were used. It should be mentioned that the listed size values are estimated rounded values and the derived *H*-values contain a corresponding uncertainty.

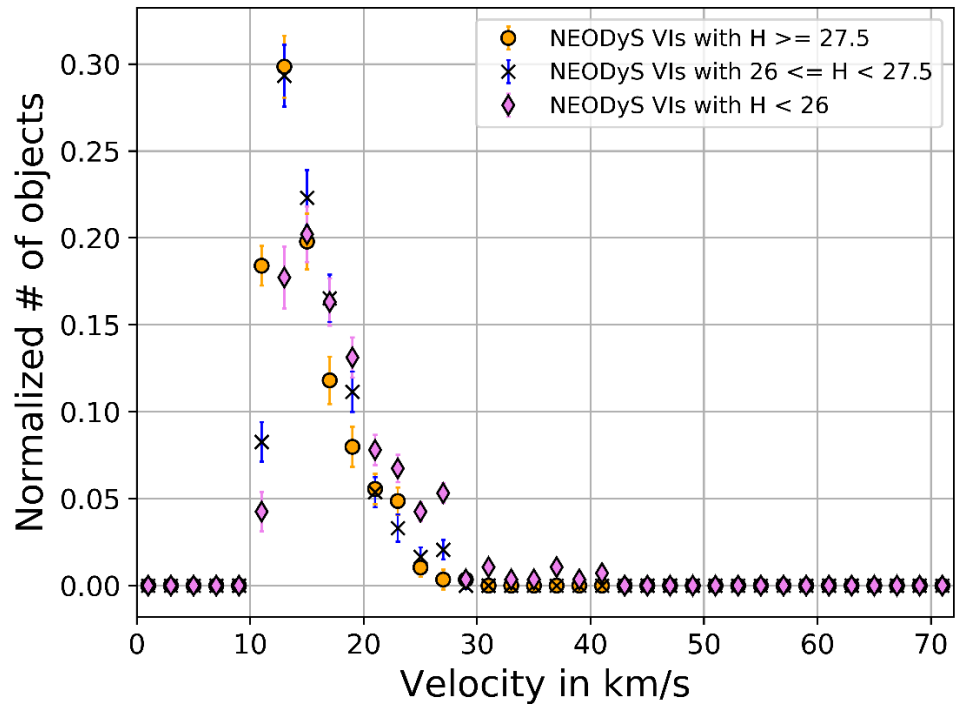

*Figure 3: Velocity distribution of NEODyS Virtual Impactors (NEODyS VIs) subsets, separated depending on the objects' absolute magnitude values.*
*The plot shows the normalized velocity distribution of the objects listed in the risk list of ESA's NEOCC with a corresponding absolute magnitude of H < 26 (pink diamonds), 26 ≤ H < 27.5 (blue crosses), and H ≥ 27.5 (orange dots).*

The *H*-values of the 13 objects calculated from the size according to Eq. (4) and obtained directly from other sources show deviations of 1.3 mag on average. The maximal difference is 4.2 mag and the minimal difference 0.25 mag. For 12 of the 13 objects the difference is up to 1.4 mag, the maximal deviation of 4.2 mag for asteroid 2010 GM23 demonstrates the uncertainty of the true size and clearly favours use of the absolute magnitude for classifications.

We separated the data into subsets of asteroids with $H < 26$, $26 \leq H < 27.5$, and $H \geq 27.5$. For comparison they are displayed in one plot, see Figure 3. There are 282, 242, and 288 objects in the subsets, respectively. The subset with only the brightest objects (pink diamonds in Figure 3) has the smallest peak, which is additionally shifted by one bin (of 2 km/s) towards larger velocities, compared to the other two subsets. But in general, the different velocity distributions show no significant difference between the distributions of the NEODyS VIs in the different *H*-ranges. No significant velocity bias for certain sizes is found. Hence, the complete data set will be used in the following.

### 2.3. CILBO meteors

The Canary Island Long-Baseline Observatory (CILBO) is a camera setup on the Canary Islands consisting of five image-intensified video cameras. It is a double station setup, one station is located on Tenerife and the second on La Palma. Every station hosts two intensified CCD Cameras, with Field of Views (FOVs) of 22° x 28° and 60° x 60°. On Tenerife a further camera is located for meteor spectra recordings. For this study the data collected by the cameras with the smaller FOV are utilized. They are called ICC7 (Tenerife) and ICC9 (La Palma). They are pointing at a spot in the sky between the two islands at a height of roughly 100 km above the surface of the Earth, covering the same volume of atmosphere. This way CILBO is able to deliver stereoscopic meteor observation data. For further information about the CILBO system see Koschny *et al*. (2013). The meteor detection is carried out with MetRec (Molau, 1999). The software provides, amongst others, the meteor's position in celestial coordinates and its apparent magnitude. Meteors that were detected with both cameras simultaneously were analysed with the Meteor Orbit and Trajectory Software (MOTS, Koschny and Díaz del Río, 2002). More recently, Albin *et al*. (2015) developed a Monte-Carlo based extension. A further enhancement of the data processing was done by Schmidt (2019), in particular studying the velocity measurement errors.

For a double station meteor, the mass of the corresponding meteoroid $m$ can be computed by considering the meteoroids mass relation on the brightness and velocity of the corresponding meteor. For this data set a formula published by Verniani (1973) was used, see Eq. (5), with the velocity $v$ of the meteor and its absolute magnitude at peak brightness $M$ computed by MOTS. The formula is based on Verniani's analysis of data by the Harvard Photographic Meteor Project (described in Verniani (1965)) as well as of radio meteor data (Verniani, 1973). It has to be mentioned that Eq. (5) uses cgs units.

$$m = 10^{\frac{-M+64.09-10\cdot\log(v)}{2.5}} \qquad (5)$$

For the double station meteors, the flux density of the corresponding meteoroids can be derived as described in detail in Koschny *et al.* (2017). They corrected the data for observational biases, like the velocity bias. This was done by assuming that for every mass bin the velocity distribution is the same. For every mass bin the observed velocity distribution was compared to a theoretical distribution, for which the ECSS based data was used, compare Section 2.4. This way, a de-biasing factor could be determined for the corresponding mass bin and the observed data could be corrected for the velocity bias, and hence for the meteors that could not be detected since they were too slow and hence, too faint.

One approach to reduce biases is to exclude all small meteoroids and to only use the ones that are large enough to have been detected by the CILBO system regardless of velocity. Using only these meteoroids should result in a velocity distribution without the bias towards high velocities. Following Koschny *et al.* (2017) all meteoroids in the order of 1 g can be detected with the CILBO system. Hence, all meteoroids with a mass equal to or larger than 1 g are used. This would mean, even with the lowest threshold velocity of 11.1 km/s a meteoroid with a mass of 1 g would cause a bright enough meteor to be detected (about 3.6 mag following the Verniani formula).

CILBO has been active since 2011 and the data set used in this work was collected between 13 September 2011 and 27 January 2018. The two CILBO stations record a common area between the islands of La Palma and Tenerife. In the time frame from Sep 2011 to Jan 2018 the cameras operated simultaneously for about 6560 hours. In this period, the two cameras observed 42957 and 55253 meteors, respectively. 18693 meteors were identified as simultaneous observations, i.e. about 30 to 40 % of the events from one station. This can be explained by the fact that meteors which are too close to or too far from one camera will not be in the field of view of the other camera. For 8811 meteors out of the 18693, good orbits have been computed. The main reason for this difference is the fact that we have excluded meteors which were only detected on three video frames, since for these only a lower limit for the velocity can be given. Some other exclusions come from the fact that the uncertainties in the astrometry were too large for the orbit algorithm to converge, e.g. for meteors close to the edge of the field of view of one camera. Only 310 of those 8811 meteoroids have a mass of at least 1 g, according to the mass formula of Verniani (Eq. (5)), and a velocity between 10 km/s and 76 km/s (range chosen to exclude outliers and for binning purposes). The velocity distribution of the corresponding meteors in 2 km/s bins is shown as violet diamonds in Figure 4. The vertical bars give the statistical errors. The smallest velocity is 10.3 km/s and the fastest 37.2 km/s. The mean velocity is about 18 km/s and the median 16.7 km/s. It has to be noticed that the velocity and magnitude values used in this work are the ones computed by MOTS using ICC9 as baseline during the computations with MOTS. Using ICC7 as baseline would result in slightly different results but a similar velocity distribution. An investigation of this difference in the data is planned as future work. The CILBO data shown include meteoroids from streams. Stream particles typically have rather high velocities. This is indicated by the ochre dashed-dotted vertical bars in Figure 4 which mark the velocities of the 12 major annual meteor streams. In our considered CILBO data set of larger objects we find few meteors with impact parameters consistent with stream particles. In a different study it was shown by Suggs *et al.* (2014) that the peak rates of the lunar impact flashes show a clear correlation with the meteor showers. Their study included meteoroids with masses of at least a few tens of grams. This could be another case of a bias towards higher velocity impactors. Meteoroids with the typical impact velocity of sporadics will require a higher mass to produce a visible impact flash.

The velocity uncertainty of the CILBO meteor data is believed to be 1 km/s - 2 km/s for the larger objects of interest in this study. It should be pointed out that various rather different equations exist to convert the measured brightness and velocity of a meteor to the mass of the source meteoroid (see e.g. Koschny *et al*. (2017)). For this study Eq. (5) developed by Verniani (1973) was used. The choice of the relationship between meteor luminosity and the mass and velocity of the source meteoroid (luminosity function) does determine how many meteors with a given speed fall above a specified mass threshold. However, in this study we only want to assure that the CILBO data considered are reasonably free of velocity bias. The meteors which are believed to be detectable by the CILBO cameras at all velocities should lie well above the selected mass threshold. For the luminosity function from Verniani our threshold corresponds to a mass of 1 g. This mass threshold would be different for a different luminosity function. Our results should be for objects big enough to be recorded by CILBO even at the lowest possible entry velocities. The 1 g mass limit is just an indication of this threshold for the CILBO capability. We do not claim that the presented speed distribution is valid for any range of smaller sizes.

It is also possible to fit a logarithmically spread normal distribution to the CILBO data. This distribution was found to yield good results for the SSA and JPL data and a Kolmogorov–Smirnov test shows that at a significance level of 0.01 the fit and the data points do not differ significantly, for the CILBO data. The function shown in Eq. (6) is the distribution presented in Figure 5 as a violet dashed-dotted line. The mean value and standard deviation of the distribution are $\mu_{CILBO\ meteors} = 16.75$ km/s with $\sigma_{CILBO\ meteors} = 4.20$ km/s.

$$y = 0.0024 + \frac{1.83}{\sqrt{2\cdot\pi}\ \cdot 0.25 \cdot x} \cdot exp\left(-\frac{\left(ln\left(\frac{x}{16.25}\right)\right)^2}{2\cdot 0.25^2}\right) \quad (6)$$

Our presented velocity distribution is in good agreement with published distributions. Most meteors are in the range of low speeds. The overall shape of the velocity distribution is in good agreement with e.g. the one of Chesley and Spahr (2004), even if both are based on very different data sets. The Chesley and Spahr impactors are based on simulated impactors based on the known NEO population, the distribution presented in this chapter is based on camera data of detected meteors. The mean of the lognormal fit of the speed distribution of the synthetic impactors by Chesley and Spahr is 16.6 km/s as published by Rumpf *et al.* (2017). This is in very good agreement with our value of $\mu_{CILBO\ meteors} = 16.75$ km/s.

### 2.4. ECSS for 100 km height

A theoretical velocity distribution is computed based on the ECSS (European Cooperation for Space Standardization, ECSS, 2008) normalized velocity distribution for the flux model of sporadic meteoroids in 1 au distance to Earth, and hence in free space. The reference model in the standard is developed by Taylor (1995) using the data of the Harvard Radio Meteor Project (HRMP) and modified to take the Earth's gravitational attraction into account. The procedure to do so is described in the standard and was used by Drolshagen *et al.* (2014) to compute the velocity distribution of the sporadic meteoroid flux in 100 km height above the Earth's surface. For every velocity value $v$ it is possible to compute the changes of the velocity due to Earth's gravity at certain distances of the meteoroids to the Earth. These changes $G$ can be computed with Eq. (7) - Eq. (9), whereas $v_\infty$ indicates the velocity in free space and $v_{esc}$ the escape velocity for a given distance of the meteoroid to the centre of the Earth, $r$. $\mu = 3.986 \cdot 10^5\ km^3/s^2$ is the result of multiplying the Earth's mass with the constant of gravity. After applying the velocity changes the adapted velocity distribution can be computed. The procedure is explained in more detail in Drolshagen *et al.* (2014). For a normalization the data is binned in 2 km/s bins, according to the different data sets analysed in this work. The resulting curve is presented as a solid blue line in Figure 4. The curve shows a rather sharp peak around 13 km/s. This is partly an artefact of the calculation and binning procedure. For infinitesimal small bins the curve will diverge at the escape velocity. Depending on the selected bin size and sampling procedure the peak will appear more or less sharp and could move slightly in position (Drolshagen *et al.*, 2014). However, the basic feature of the velocity distribution with a maximum just above the escape velocity does not depend on the calculation procedure.

$$G = \frac{v^2}{v^2 - v_{esc}^2} \tag{7}$$

$$v^2 = v_{esc}^2 + v_\infty^2 \tag{8}$$

$$v_{esc}^2 = \frac{2 \cdot \mu}{r} \tag{9}$$

### 2.5. Major meteor showers

Meteor showers cause a lot of attention and are shifting our perception of meteors. But even though this would result in a bias towards higher velocities, this bias is not included in the data we present in this work, as it can be clearly seen in Figure 4.

The vertical ochre dashed-dotted lines in the figure represent the 12 major meteor showers. These are included for comparison to show the typical higher velocities of shower meteors. The height of the lines has no meaning. Meteor showers happen when the Earth crosses a meteoroid stream in our solar system and all meteoroids have a very similar velocity since they originate from the same parent body. Consequently, the meteors of one shower all have approximately the same velocity. Since the meteoroid streams can be distributed unevenly along their orbit around the sun, the intensity of the showers can vary from year to year. Furthermore, the showers are active during the same range of dates every year. Every year, the International Meteor Organisation (IMO) publishes a shower calendar including amongst others the activity range of dates, the date of maximum activity, the velocity of the shower in free space relative to Earth, and its ZHR (Zenithal Hourly Rate) (IMO, 2017). The ZHR gives the number of shower meteors an observer would see per hour if the radiant of the shower would be at the zenith and his limiting magnitude was 6.5 mag (Brown, 1990). This value for the showers' activity in 2018 was used to choose the major showers listed in Table 1, all of them have a predicted ZHR $\geq$ 10 in 2018. These showers are marked in Figure 4 with ochre dashed-dotted lines. It has to be mentioned that two times two showers have the same velocity: The Quadrantids and the South $\delta$-Aquariids both have a velocity of 41 km/s and the $\eta$-Aquariids and the Orionids have a velocity of 66 km/s. The lack of showers with smaller velocities indicates that it is very likely that the distribution of objects at the lower velocities are comprised of sporadics. Furthermore, it could indicate that the subjective velocities of meteors are rated unrealistic high, as meteor streams receive increased attention and are closely monitored.

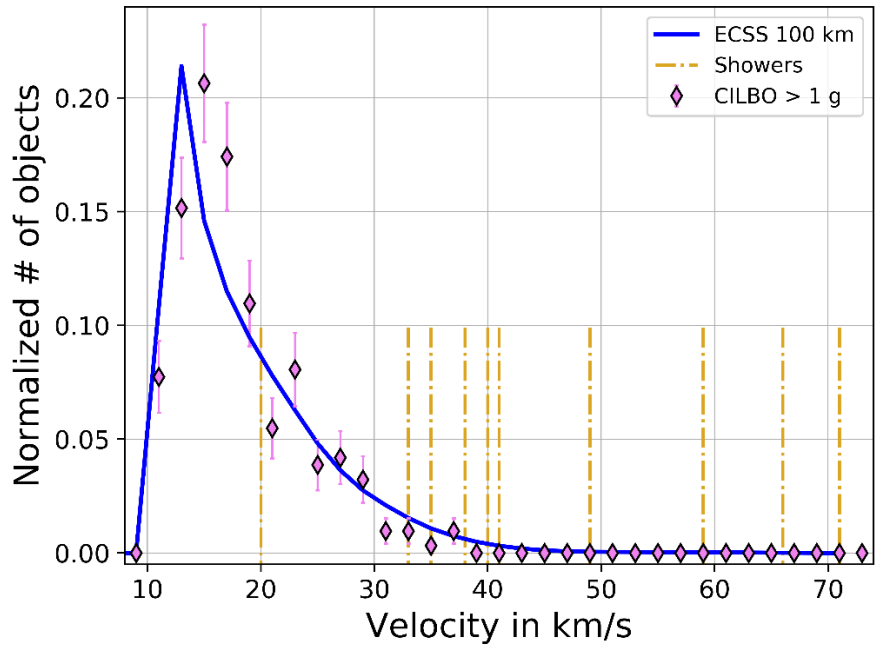

*Figure 4: The blue solid line represents the velocity distribution, adapted to 100 km height from the ECSS model, as described in Section 2.4. The violet diamonds represent the velocity distribution of the larger objects for which detection with the CILBO double-camera system should be complete. The vertical bars give the statistical errors, as described in Section 2.3. Furthermore, the vertical ochre dashed-dotted lines indicate the velocities of major meteor showers as introduced in Section 2.5, see Table 1. The height of the meteor shower bars has no meaning.*

*Table 1: Meteor showers with a ZHR ≥ 10 in 2018 following the IMO's shower calendar. The name of the shower is listed sorted by its date of maximum activity (column 3). The shower's velocity is in column 4 and its ZHR in the last column (IMO, 2017).*

| Shower | IAU Code | Maximum Activity Date | Velocity km/s | ZHR |
| --- | --- | --- | --- | --- |
| Quadrantids | 010 QUA | 03 Jan | 41 | 110 |
| Lyrids | 006 LYR | 22 Apr | 49 | 18 |
| $\eta$-Aquariids | 031 ETA | 06 May | 66 | 50 |
| Daytime Arietids | 171 ARI | 07 Jun | 38 | 30 |
| South $\delta$-Aquariids | 005 SDA | 30 Jul | 41 | 25 |
| Perseids | 007 PER | 12 Aug | 59 | 110 |
| Draconids | 009 DRA | 09 Oct | 20 | 10 |
| Orionids | 008 ORI | 21 Oct | 66 | 20 |
| Leonids | 013 LEO | 17 Nov | 71 | 15 |
| Puppid-Velids | 301 PUP | 07 Dec | 40 | 10 |
| Geminids | 004 GEM | 14 Dec | 35 | 120 |
| Ursids | 015 URS | 22 Dec | 33 | 10 |

### 2.6. Comparison

The previous sections introduced different data sets, providing information on the velocity of objects in space. All velocity distributions are normalized and presented in Figure 5. The velocity distribution, adapted to 100 km height from the data of the ECSS (blue solid line), shows a clear maximum around 13 km/s. As expected, there are no values lower than 11 km/s, the escape velocity. Roughly 68 % of the area under the curve is lower than 20 km/s and 93 % lower than 30 km/s. The distribution shows that most objects have a velocity in the lower range of the possible velocities and that there are only a few fast objects.

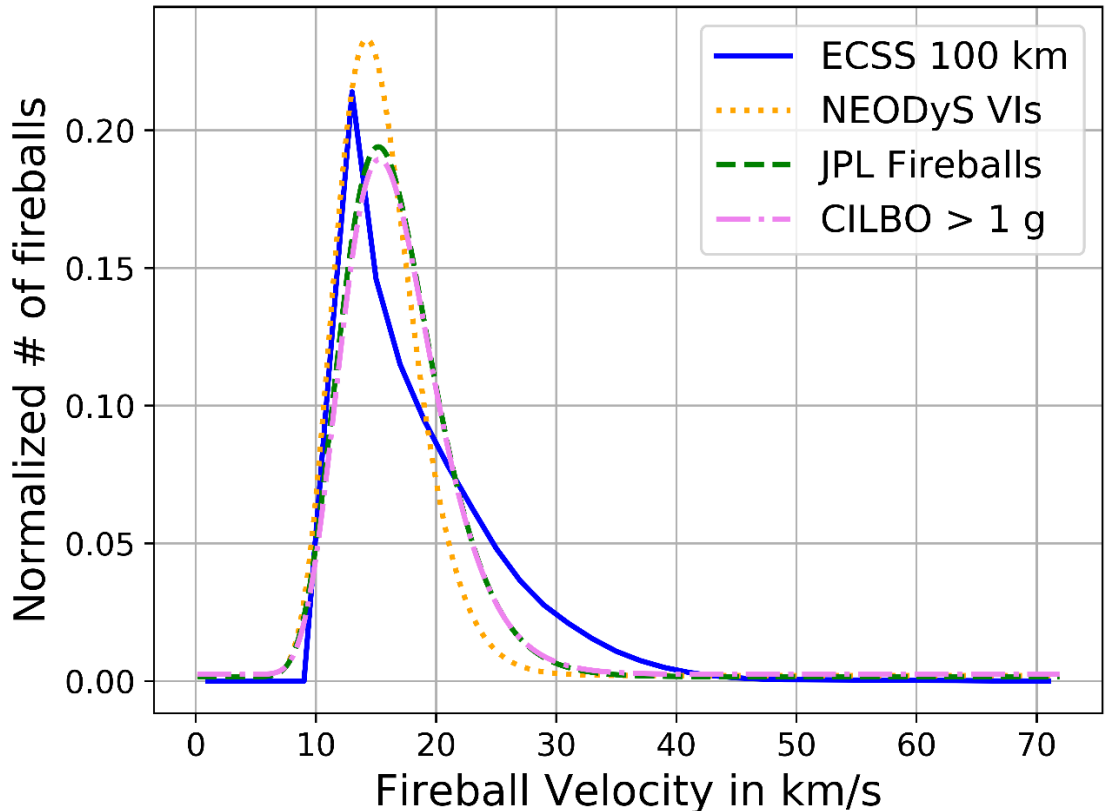

*Figure 5: Earth impact velocity distributions of larger objects in 2 km/s bins. The blue solid line represents the velocity distribution, adapted to 100 km height from the data of the ECSS, as described in Section 2.4. The green dashed line shows the fit to the velocity distribution of the objects in the fireball database published by NASA's CNEOS JPL, compare Section 2.1. The orange square-dotted line describes the fit adapted to the velocity distribution of the NEODyS VIs, the objects listed in the risk list of NEAs of ESA's SSA, see Section 2.2. The violet dashed-dotted line represents the velocity distribution of the larger objects detected with the CILBO double-camera system, as described in Section 2.3.*

The velocity distribution of the objects which are listed in the risk list of ESA's SSA (NEODyS VIs) (represented by the fit as a orange square-dotted line) shows a clear peak at low velocities, too. The maximum is at around 14 km/s and the distribution shows a similar behavior as the ECSS reference model. The velocity distribution of the objects in the fireball database published by NASA's CNEOS at JPL represented by the fit (green dashed line) has its highest value at around 15 km/s. The JPL and NEODyS VI data both show a wider peak and a steeper decrease for velocities above around 20 km/s. The shift of the JPL data towards higher velocities could be due to the larger scattering of the data points (compare Figure 1), which could possibly be explained by the lower number of objects in the data set for which the velocity is given. The velocity distribution of the larger objects detected with the CILBO double-camera system (violet dashed-dotted line) has the same shape with a clear maximum at 15 km/s. The maximum of the curve is very similar to the JPL distribution. To conclude, all data sets show a very similar velocity distribution.

In contrast, the meteors corresponding to major showers (vertical ochre dashed-dotted lines in Figure 4) show higher velocities.

The different data sets determined the impact velocities at different heights. The SSA give its values for the Earth's surface, neglecting the atmosphere (Faggioli, 2019). The ECSS data is adapted from free space to the top of the atmosphere, at 100 km height above the Earth's surface (Drolshagen *et al*., 2014). The velocities of the fireballs in the JPL database are given for the individual location before the impact explosion, ranging from 15 km to 74 km with a mean of about 36 km (CNEOS, 2018). Most smaller meteors occur at a height of about 80 km – 100 km above the Earth's surface. Larger asteroids penetrate deeper into the atmosphere before they reach their maximum brightness (Borovička *et al.*, 2015). The CILBO meteors taken into account for this work occur mostly at a height of about 90 km, ranging from 75 km to 127 km with an average height of 89 km. If and how the differences in height influence the object's velocity will be discussed briefly.

To highlight the differences and the possible implications, one event is discussed in more detail, the asteroid 2008 $TC_3$. 2008 $TC_3$ was the first asteroid that has been detected by telescope before its entry into the Earth's Atmosphere. It was discovered on 06 October 2008 at 06:39 UTC by the Catalina Sky Survey telescope and entered the Earth's Atmosphere over Sudan on 07 October 2008 at 01:49 UTC. The entering asteroid had a pre-atmospheric mass of about 83 t from which some survived and was partially recovered as meteorites (Jenniskens *et al.,* 2009). For further information about the 2008 $TC_3$-event over Sudan see e.g. Jenniskens *et al.* (2009). Till today, as far as is known, there are only three asteroids which have been detected before their entry.

The 2008 TC$_3$ event is listed in the JPL database. They give the date for the impact to be 07 October 2008 at 02:45 UTC and a location at 20.9° N and 31.4° E. The altitude of the fireball is listed as 38.9 km, its velocity as $v_{JPL} = 13.3$ km/s and the calculated total impact energy as $1$ kt TNT equivalent (CNEOS, 2018).

The impact velocity of the asteroid at the top of the atmosphere, at 100 km height above the Earth's surface and obtained from the orbital data, is: $v_{orbit} = 12.78$ km/s (Micheli, 2019) The orbital velocity data as well as the JPL velocity data are given in geocentric Earth-fixed coordinates but differ by more than 0.5 km/s.

The changes in velocity due to the Earth's attraction can be described with the Eq. (7) – (9). From those it is clear that the escape velocity for the different heights can be easily compared to get a feeling for the respective changes. Hence, the escape velocity for 100 km above the Earth's surface and at 38.9 km (the JPL height value for 2008 TC$_3$) were computed. They are $v_{esc,\ 100\ km} = 11.1$ km/s and $v_{esc,\ 38.9\ km} = 11.15$ km/s, respectively. This small velocity increase is probably at least compensated by atmospheric friction. A quantitative assessment on the atmospheric impact on the velocities would go beyond the scope of this work but we do not expect a large effect on the velocities by atmospheric friction. Furthermore, the atmospheric friction would at most decelerate the asteroid, but the JPL velocity of the meteoroid when it has passed more atmosphere is higher.

For this reason, we expect the differences in the velocity determinations to result from uncertainties in the measurements. To conclude, the differences in height for which the velocities in the different databases are given are expected to have a negligible impact on the velocity distributions.

3. **Conclusion**

Different velocity distributions of large meteoroids and asteroids impacting Earth were analysed and compared in this work based on different data sets. The aim was to obtain the unbiased velocity distribution of large meteoroids or asteroids which actually have impacted Earth (JPL fireball and CILBO data) or which are based on well-defined impact solutions of known asteroids (NEODyS VIs). The mean values of the impact velocities of the objects in the different data sets were found to be $v_{mean,\ JPL\ fireballs} = 17.73$ km/s, $v_{mean,\ NEODyS\ VIs} = 16.48$ km/s, and $v_{mean,\ CILBO\ meteors} = 18.04$ km/s and the median values $v_{median, JPL\ fireballs} = 16.70$ km/s, $v_{median, NEODyS\ VIs} = 15.33$ km/s, and $v_{median, CILBO\ meteors} = 16.68$ km/s.

The obtained velocity distributions agree very well with the distribution of the ECSS, which is based on the HRMP results, often used as the standard reference model. For the considered distributions of Earth impacting objects, which should be unbiased by velocity it was found that most have a relative slow impact velocity just above the escape velocity. The different distributions have their maxima at around $v_{max,\ NEODyS\ VIs} = 14$ km/s, $v_{max,\ JPL\ fireballs} = 15$ km/s, and $v_{max,\ CILBO\ meteors} = 15$ km/s. This compares well with the maximum of the standard velocity distribution assumed for meteoroids of $v_{max,\ ECSS} = 13$ km/s.

The basic distribution looks very similar to the expected velocity distribution of Earth impactor speeds from various de-biased NEA models (Chesley and Spahr, 2004, Greenstreet *et al.*, 2012, and Le Feuvre and Wieczorek, 2011).

The velocity distributions can be described with a logarithmically spread normal distribution. The results show mean values (expected values) and standard deviations of the distributions of $\mu_{JPL\ fireballs} = 16.69$ km/s with $\sigma_{JPL\ fireballs} = 4.25$ km/s, $\mu_{NEODyS\ VIs} = 15.31$ km/s with $\sigma_{NEODyS\ VIs} = 3.43$ km/s, and $\mu_{CILBO\ meteors} = 16.75$ km/s with $\sigma_{CILBO\,meteors} = 4.20 \text{ km/s}$. This is in good agreement with the results of Rumpf *et al.* (2017) and their value of $\mu_{Rumpf\,et\,al.}$ = 16.6 km/s.

The three data sets analysed in this work are all for objects impacting Earth but they cover different size ranges. The NEODyS data are for known asteroids in space which span a size range of a few metres to several hundred metres. The fireball data from the JPL database are for objects between about 1 m in size and a maximum of ca 20 m (the Chelyabinsk object). The CILBO data are for meteoroids large enough to be recorded by the video cameras at all velocities (ca larger than 1 g in mass). All three data sets should be unbiased what concerns the impact velocity.

The velocity distributions for all three sets of data are remarkably similar. They have their maxima at relative low velocities. This agrees with most other meteor velocity distributions recorded optically, by radar or radio after de-biasing. These distributions usually peak at higher velocities as the signal created by a given object is of a strong function of velocity. At higher velocities more abundant smaller particles become visible. The Apex contribution and the dominance of meteor showers are manifestations of this velocity bias. This well-known velocity bias has to be carefully considered if fluxes as function of mass are derived.

Only few objects with high velocities are present in the unbiased distribution presented here. The larger objects in the JPL and NEODyS data sets should be of asteroidal origin. Recent meteoroid population models assume that comets are the clearly dominating source for meteoroids up to several hundred grams in mass (McNamara *et al.*, 2004, Dikarev *et al.*, 2005). That implies that the CILBO objects > 1 g could be mainly of cometary origin. But their velocity distribution shows no signs of high velocities as known from meteor stream particles. If they originate from comets, their orbits must have evolved towards more circular orbits, co-rotating with Earth as they dispersed into the sporadic background.

The present results apply to objects which have at least ca 1 g mass or (much) more. It is possible that smaller objects have a different velocity distribution. But it is also probable that observed differences are mainly an effect of the velocity bias. It should be possible to assess and correct this bias when the luminosity of a meteor as function of the source object velocity is understood in detail. Differently shaped velocity distributions for smaller objects detected by different radar systems are e.g. published by Taylor and Elford (1998), Galligan and Baggaley (2004), or Janches *et al.* (2003). Nonetheless, the more recent works by Moorhead *et al.* (2017) and Williams *et al.* (2017) are also based on radar data and presented velocity distributions similar to the ones derived in this work and the ones in literature based on de-biased NEA models.

The median value of the velocity distribution will generally give the best representation by a single velocity. The maxima will neglect the higher velocity tail of the distribution and the mean values could give too much weight to the higher velocities. The median values obtained from the data are 15.3 km/s for the asteroids in the risk list, and 16.7 km/s for both the JPL fireball data and the larger CILBO meteors, matching the expected values derived from the log-normal distributions. The latter two data sets are for objects which actually did impact Earth. Therefore, for larger meteoroids and small asteroids impacting Earth we would expect typical impact velocities around 16.7 km/s. For Earth orbiting spacecraft the impact velocity on a given surface will depend on the impact geometry but, because of the spacecraft velocity, the average impact velocity will be slightly higher than for the Earth's Atmosphere at around 18 km/s. If impact effects or the impact risk to spacecraft is analysed a fixed representative velocity will depend on the actual effect considered. Different effects will have different dependencies on the impact velocity and should be represented by different impact velocities. Using the full velocity distribution will overcome this problem.

4. **Acknowledgements**

We thank the European Space Agency and the University of Oldenburg for funding this work which was carried out as part of the NEMO project. Furthermore, we would like to thank Marco Micheli for its computations of the impact velocities of 2008 TC$_3$ and 2018 LA from the orbit.

**5. References**

Aerospace (2018). Reentries [online]. Available at: https://aerospace.org/reentries [Accessed: 25 Nov. 2018].

Borovička, J., Spurný, P., Brown, P., (2015). Small Near-Earth Asteroids as a Source of Meteorites, chapter, Asteroids IV, published by University of Arizona Press, 2015, Vol. 257.

Brown (1990). On the Cause and Nature of Error in Zenithal Hourly Rates, WGN, the Journal of the IMO 18:4 (1990).

Brown, P., Spalding, R. E., ReVelle, D. O., Tagliaferri, E., and Worden, S. P. (2002). The flux of small near-Earth objects colliding with the Earth. Nature, 420(Nov.), 294-296.

Brown, P., Wong, D. K., Weryk, R. J., and Wiegert, P. (2010). A meteoroid stream survey using the Canadian Meteor Orbit Radar. II: Identification of minor showers using a 3D wavelet transform. Icarus, 207(May), 66-81.

Brown, P., Wiegert, P., Clark, D., Tagliaferri, E. (2016). Orbital and physical characteristics of meter-scale impactors from airburst observations. Icarus, Volume 266, Pages 96-111, https://doi.org/10.1016/j.icarus.2015.11.022.

Ceplecha, Z., McCrosky, R.E., (1976). Fireball end heights: a diagnostic for the structure of meteoric material. J. Geophys. Res. 81/35, 6257–6275.

Ceplecha, Z. (1992). Influx of interplanetary bodies onto Earth, Astronomy and Astrophysics, 263/1-2, 361-366.

Chesley, S. R., and Spahr, T. B., (2004). Earth impactors: orbital characteristics and warning times. In Mitigation of Hazardous Comets and Asteroids, edited by Belton M. J. S., Morgan T. H., Samarasinha N. H., and Yeomans D. K. Cambridge: Cambridge University Press. pp. 22–37.

CNEOS/JPL, NASA (2018). Fireballs [online]. Available at: https://cneos.jpl.nasa.gov/fireballs/ [Accessed: 29 Jan. 2019].

Cooke, W. J., and Moser, D. E. (2010). The 2011 Draconid Shower Risk to Earth-Orbiting Satellites. NASA Technical Reports Server Document 20100024125.

Devillepoix H. A. R., Bland, P. A., Sansom, E. K., Towner, M. C., Cupák, M., Howie, R. M., Hartig, B. A. D., Jansen-Sturgeon, T., Cox, M. A. (2019). Observation of metre-scale impactors by the Desert Fireball Network. Monthly Notices of the Royal Astronomical Society, Volume 483, Issue 4, March 2019, Pages 5166–5178, https://doi.org/10.1093/mnras/sty3442

Dikarev, V., Grün, E., Baggaley, J., Galligan, D., Landgraf, M., Jehn, R. (2005). The new ESA meteoroid model. Advances in Space Research, 35, 1282-1289.

DISCOS (2018). ESA Space Debris [online]. Available at: https://discosweb.esoc.esa.int [Accessed: 20 Nov. 2018].

Divine, N. (1993). Five populations of interplanetary meteoroids. Journal of Geophysical Research, 98(Sept.), 17029-17048.

Drolshagen, G., Dikarev, V., Landgraf, M., Krag, H., and Kuiper, W. (2008). Comparison of Meteoroid Flux Models for Near-Earth Space. Earth Moon and Planets, 102(June), 191-197.

Drolshagen, E., Ott, T., Koschny, D., Drolshagen, G., and Poppe, B. (2014). Meteor velocity distribution from CILBO double station video camera data, Proceedings of the International Meteor Conference, Giron, France, pp. 16-23, ISBN 978-2-87355-028-8.

Drolshagen, G., Koschny, D., Drolshagen, S., Kretschmer, J., and Poppe, B. (2017). Mass accumulation of Earth from interplanetary dust, meteoroids, asteroids and comets. Planetary and Space Science, 143(Sept.), 21-27.

ECSS (2008). European Cooperation for Space Standardization, Space Engineering, Space Environment, ECSS-E-ST-10-04C. Noordwijk, Netherlands: ESA Requirements and Standards Division.

Erickson, J. E. (1968). Velocity Distribution of Sporadic Photographic Meteors. Journal of Geophysical Research 73:3721–3726

Faggioli, F., (2019), pers. comm.

Farnocchia, D., Chesley, S. R., Milani, A., Gronchi, G. F., Chodas, P. W. (2015). Orbits, long-term predictions, impact monitoring. In: Asteroids IV, Michel, P., DeMeo, F., and Bottke, W. F. (eds.), University of Arizona Press, Tucson, 895 pp. ISBN: 978-0-816-53213-1, 815-834.

Galligan D. P., and Baggaley W. J. (2004). The orbital distribution of radar-detected meteoroids of the Solar system dust cloud. Monthly Notices of the Royal Astronomical Society.

Greenstreet S., Ngo H., and Gladman B. (2012). The orbital distribution of Near-Earth Objects inside Earth's orbit. Icarus 217:355–366.

Grün, E., Zook, H. A., Fechtig, H., and Giese, R. H. (1985). Collisional balance of the meteoritic complex. Icarus, 62(May), 244-272.

Halliday, I., Grün, A. A., and Blackwell, A. T. (1996). Detailed data for 259 fireballs from the Canadian camera network and inferences concerning the influx of large meteoroids. Meteoritics and Planetary Science, 31(Mar.), 185-217.

Hunt, S. M., Oppenheim, M. M., Close, S., Brown, P. G., McKeen, F., and Minardi, M., 2004. Determination of the meteoroid velocity distribution at the Earth using high-gain radar. Icarus 168:34–42.

IMO (2017). International Meteor Organisation – 2018 Meteor Shower Calendar, edited by J. Rendtel.

Janches D., Nolan M. C., Meisel D. D., Mathews J. D., Zhou Q. H., and Moser D. E. 2003. On the geocentric micrometeor velocity distribution. Journal of Geophysical Research 108:101029/.

Jenniskens, P. (1994). Meteor stream activity I. The annual streams. Astronomy & Astrophysics, 287(July), 990-1013.

Jenniskens, P., Shaddad, M. H., Numan, D., Elsir, S., Kudoda, A. M., Zolensky, M. E., Le, L.; Robinson, G. A., Friedrich, J. M., Rumble, D., Steele, A., Chesley, S. R., Fitzsimmons, A., Duddy, S., Hsieh, H. H., Ramsay, G.,Brown, P. G., Edwards, W. N., Tagliaferri, E., Boslough, M. B., Spalding, R. E., Dantowitz, R., Kozubal, M., Pravec, P., Borovicka, J., Charvat, Z., Vaubaillon, J., Kuiper, J., Albers, J., Bishop, J. L., Mancinelli, R. L., Sandford, S. A., Milam, S. N., Nuevo, M., Worden, S. P. (2009). The impact and recovery of asteroid 2008 TC3, Nature 458(7237):485–488.

Jenniskens, P., Nénon, Q., Albers, J., Gural, P. S., Haberman, B., Holman, D., Morales, R., Grigsby, B. J., Samuels, D., Johannink, C., (2016). The established meteor showers as observed by CAMS. Icarus, 266(Mar.), 331-354.

Kessler, D. J. (1972). A Guide to Using Meteoroid-environment Models for Experiment and Spacecraft Design Applications. NASA/TN-D-6596, Mar.

Koschny, D., and Díaz del Río, J. (2002). Meteor Orbit and Trajectory Software (MOTS) - Determing the Position of a Meteor with Respect to the Earth Using Data Collected with the Software MetRec. WGN, the Journal of the IMO 30:4, pp. 87-101.

Koschny, D., Bettonvil, F., Licandro, J., van der Luijt, C., Mc Auliffe, J., Smit, H., Svendhem, H., de Wit, F., Witasse, O., Zender, J., (2013). A double-station meteor camera setup in the Canary Islands - CILBO. Geoscientific Instrumentation Methods and Data Systems, 339-348.

Koschny, D. and Drolshagen, G., (2015). Activities in Europe related to the mitigation of the threat from near-Earth objects, Advances in Space Research, Volume 56, Issue 3, pp. 549-556.

Koschny, D., Drolshagen, E., Drolshagen, S., Kretschmer, J., Ott, T., Drolshagen, G., and Poppe, B. (2017). Flux densities of meteoroids derived from optical double-station observations. Planetary and Space Science, 143(Sept.), 230-237.

Laplace, P.-S. (1774). Mémoire sur la probabilité des causes par les évènements. Mémoires de l'Academie Royale des Sciences Presentés par Divers Savan, 6, 621–656.

Le Feuvre, M., Wieczorek, M.A. (2011). Nonuniform cratering of the Moon and a revised crater chronology of the inner solar system, Icarus, doi: 10.1016/j.icarus.2011.03.010

Limpert, E., Stahel, W. A., and Abbt, M. (2001). Log-normal Distributions across the Sciences: Keys and Clues. BioScience, May 2001, Vol. 51 No. 5, pp 341-352.

Liu, X., Sachse, M., Spahn, F., and Schmidt, J. (2016). Dynamics and distribution of Jovian dust ejected from the Galilean satellites. Journal of Geophysical Research (Planets), 121(July), 1141-1173.

McNamara, H., Suggs, R., Kauffman, B., Jones, J., Cooke, W., Smith, S. (2004). Meteoroid Engineering Model (MEM): A Meteoroid Model for the Inner Solar System. Earth Moon and Planets, 95(Dec.), 123-139.

McAlister, D. (1879). The law of the geometric mean. Proceedings of the Royal Society 29: 367–376.

McWilliam, I. G. and Bolton, H. C., (1969). Instrumental peak distortion. I. Relaxation time effects, Anal. Chem., 41, 1755-1762, DOI: 10.1021/ac60282a001.

Micheli, M. (2019). pers. comm.

Molau, S. (1999). The meteor detection software MetRec. In: Arlt, R., Knöfel, A. (eds.), Proceedings of the International Meteor Conference, Stara Lesna, Slovakia, 20-23 Aug 1998, IMO, 9-16.

Moorhead, A. V., Wiegert, P. A., and Cooke, W. J. (2014). The meteoroid fluence at Mars due to Comet C/2013 A1 (Siding Spring). Icarus, 231(Mar.), 13-21.

Moorhead, A. V., Brown, P. G., Campbell-Brown, M. D., Denis Heynenb, D., Cooke, W. J. (2017). Fully correcting the meteor speed distribution for radar observing biases. Planetary and Space Science 143 (2017) 209–217

NEODyS (2018). NEODyS-2 [online]. Available at: https://newton.spacedys.com/neodys/ [Accessed: 29 Nov. 2019].

OriginLab (2018). Origin and OriginPro [online]. Available at: https://www.originlab.com/index.aspx?go=PRODUCTS/Origin [Accessed: 16 Nov. 2018].

OriginLab (2018b). LogNormal [online]. Available at: https://www.originlab.com/doc/Origin-Help/LogNormal-FitFunc [Accessed: 16 Nov. 2018].

OriginLab (2018c). Asym2Sig [online]. Available at: https://www.originlab.com/doc/Origin-Help/Asym2Sig-FitFunc [Accessed: 16 Nov. 2018].

OriginLab (2018d). GaussMod [online]. Available at: https://www.originlab.com/doc/Origin-Help/GaussMod-FitFunc [Accessed: 16 Nov. 2018].

OriginLab (2018e). Laplace [online]. Available at: https://www.originlab.com/doc/Origin-Help/Laplace-FitFunc [Accessed: 16 Nov. 2018].

Python Software Foundation (2018). Python [online]. Available at: https://www.python.org/ [Accessed: 16 Nov. 2018].

Oldenburg, 14 November 2019                                    NEMO-PA-004_1_2_velocity-distributionsRumpf C. M., Lewis H. G., and Atkinson P. M. (2017). Asteroid impact effects and their immediate hazards for human populations. Geophysical Research Letters 44:3433–3440.

Sabarinath, A., and Anilkumar, A. K. (2008). Modeling of Sunspot Numbers by a Modified Binary Mixture of Laplace Distribution Functions, Solar Phys (2008) 250: 183–197 DOI 10.1007/s11207-008-9209-5.

Schmidt, K. (2019). Analysis of uncertainties and algorithmic errors in a double-station meteor camera setup. Univ. Oldenburg, Germany, Master's Thesis, ESA-MRG-RP-045.

Sekanina, Z. and Southworth, R. B. (1975). Physical and dynamical studies of me-teors. Meteor-fragmentation and stream-distribution studies, NASA Con-tractor Report CR-2615, 94pp., Smithsonian Institution, Cambridge, Ma, USA

Smith, R. E. (1994). Natural Orbital Environment Guidelines for Use in Aerospace Vehicle Development. NASA/TM-4527, June.

SonotaCo. (2009). A meteor shower catalog based on video observations in 2007-2008. WGN, Journal of the International Meteor Organization, 37(Apr.), 55-62.

SSA (2018). Risk Page [online]. Available at: http://neo.ssa.esa.int/risk-page [Accessed: 29 Jan. 2019 with last update: 28.01.2019, 16:09 UTC].

Staubach, P., Grün, E., and Jehn, R. (1997). The meteoroid environment near Earth. Advances in Space Research, 19(May), 301-308.

Suggs, R.M., Moser, D.E., Cooke, W.J., Suggs, R.J. (2014). The flux of kilogram-sized meteoroids from lunar impact monitoring. Icarus 238, 21-36.

Taylor, A. D. (1995). The Harvard Radio Meteor Project Meteor Velocity Distribution Reappraised. Icarus, 116(July), 154-158.
39

Taylor, A. D., and Elford, W. G. (1998). Meteoroid orbital element distributions at 1 AU deduced from the Harvard Radio Meteor Project observations. Earth Planets Space,50, 569–575.

Verniani, F., (1965). On the luminous efficiency of meteors. Smithsonian Contributions to Astrophysics. 8, 141–171.

Verniani, F., (1973). An analysis of the physical parameters of 5759 faint radio meteors. J. Geophys. Res. 78, 8429–8462.

Weryk, R. J. and Brown, P. G., (2013). Simultaneous radar and video meteors - II: Photometry and ionisation. Planetary and Space Science 81, 32–47.

Williams E. R., Wu Y. J., Chau J. L., and Hsu R. R. (2017). Intercomparison of radar meteor velocity corrections using different ionization coefficients. Geophysical Research Letters 44:5766–5773.

Xun, Z., Tang, G., Song, L., Han, K., Xia, H., Hao, D., and Yang, Y. (2014). Dynamic scaling behaviours of the Etching model on fractal substrates, J. Stat. Mech. (2014) P12008, doi:10.1088/1742-5468/2014/12/P12008.